\documentclass{article}
\usepackage[utf8]{inputenc}
\usepackage{url}

\title{ReviewCoin: Paying for Real Work}
\author{Chris Welty \thanks{The author has benefitted from discussions of this idea over the past five years with many colleagues, including Lora Aroyo, John Domingue, Harald Sack, Axel Polleres, Valentina Presutti, Jakub Tomczak, Andrew McCallum, and others.}\\ 
Google Research
}
\date{Dec. 2024}

\usepackage{natbib}
\usepackage{graphicx}

\begin{document}

\maketitle
\begin{abstract}
    The peer-review process is broken and the problem is getting worse, especially in AI: large conferences like NeurIPS increasingly struggle to adequately review huge numbers of paper submissions.     I propose a scalable solution that, foremost, recognizes reviewing as important, necessary, \emph{work} and rewards it with crypto-coins owned and managed by the conferences themselves.  The idea is at its core quite simple: paper submissions require work (reviews, meta-reviews, etc.) to be done, and therefore the submitter must pay for that work.  Each reviewer submits their review to be approved by some designated conference officer (e.g. PC chair,  Area Chair, etc.), and upon approval is paid a single coin for a single review.  If three reviews are required, the cost of submission should be three coins + a tax that covers payments to all the volunteers who organize the conference.  After some one-time startup costs to fairly distribute coins, the process should be relatively stable with new coins minted only when a conference grows.
\end{abstract}

\section{Disclaimer}
The ideas expressed below are solely those of the author and not his employer, and do not represent a committment by any individual or corporation to implement or support the system proposed.  The purpose of this paper is to present some ideas for helping mitigate the problems of scale experienced in the peer review system, in order to collect feedback and discuss these ideas openly.

\section{Introduction}
One could go on at length criticizing the existing peer-review system and the way it appears perpetually on the verge of collapse.  Several  solutions to various problems with peer review have been proposed and deployed, most notably \emph{rolling-reviews}, which allow papers to be tracked across participating conferences, potentially reducing the work of follow-up reviews, and \emph{open reviews}, which obviate the need for rolling reviews by making the previous reviews public.

No solution proposed so far seems to address the imbalance between the scale of submissions and the scale of reviews required to deal with them.  In general, a fair system should recognize that any submission has a cost -- the reviews, primarily, as well as the work that goes into turning those reviews into a decision -- and the burden of that cost should be borne by the submitter, not thrown into the community at large under some vague moral statute ("doing reviews is part of your responsibility as a researcher").  Therefore I propose a simple, in principle, extension to the conference review system that requires submitters to pay for the work of reviewing their submission using a crypto-coin that is owned and managed by the conferences themselves.

\section{Basic Idea}
Each paper submission costs the conference a certain amount of work, and this cost should be borne by the submitter.  As long as this balance is maintained, the process should scale indefinitely; to maintain the balance, periodic analysis and tweaking to the amount of available coins, the pay rate, and the tax, will be necessary.

The cost of submission is the cost of reviews plus a tax to defray the other costs required in processing reviews and turning them into decisions, extra reviews on some papers, organizing the program, and all the other "free" work that conferences utilize today.  

For simplicity we assume the cost of a review to be a unit cost in the currency: 1 review is worth 1 coin.  Therefore the cost borne by the submitter of 1 paper is $\rho$ (the number of reviews required) + $\tau$ (the tax, which is determined before submission time to defray all the other payments the conference will make).

To achieve the desired balance between reviews and papers, we maintain the ledger of debts (from authors to conferences, and conferences to reviewers) in a blockchain and introduce a crypto currency, \emph{ReviewCoin} (RC), for representing the value of reviews and the cost of submission.   

\section{Basic Process}
The process has a startup cost which is outlined later, but should stabilize quickly. At regular intervals, organizations should review the landscape and decide if new coins, representing community growth, are required.

\begin{itemize}
    \item An editor (program chair, journal editor) indicates papers may be submitted.
    
    \item Authors submit a paper and pay $\rho + \tau$ RC, where $\rho$ is the number of basic reviews and $\tau$ is a tax representing the extra costs that may occur on some papers, amortized across all the submissions.  Authors must specify a single user, e.g. the "corresponding author", whose account will be charged for the submission.  Collaborators can transfer any existing RCs they have to share the cost.
    
    \item The editor "hires" $\rho$ reviewers to review the paper.
    
    \item The reviewers submit their reviews.
    
    \item The Editor, or some designee (e.g. an Area Chair) reviews the reviews and authorizes the payment of 1 RC to the author of each approved review. 

    \item After a specified interval, submissions are closed, with $n$ papers submitted.  After the initial reviews are paid, there will be a surplus of $\tau \times n$ RCs, which can be paid to the editor, extra reviewers when required, and to other members of the organization for their `volunteer' work.  
\end{itemize}

In the absence of any exceptions to this process, the system clearly maintains a balance between reviews and submissions.  Exceptions, of course, will happen, and the process should be robust to them:

\begin{itemize}
    \item Authors of non-approved reviews are not paid, but can revise and re-submit, with the goal of ultimately being paid.  If a conference "requires" reviewers respond to author comments, the conference now has the "teeth" to enforce that requirement: write good reviews and respond to authors or you don't get paid.
    
    \item Editors should be paid for their work as well, which should be some function of the number of papers and reviews they handle.  Multiple editors (e.g. a senior program committee) or hierarchies of editors must be cost-modeled and folded into the value of $\tau$.
    
    \item Some papers require more than $n$ reviews to reach a decision.  The rate of extra reviews should be modeled from history and also folded into the tax.
    
    \item Authors, who are now "paying" to submit, will experience heightened sensitivity to rejection.  Editors may optionally allow for an author to \emph{challenge} the decisions by reviewers.  The author must pay for the challenge, the cost is the number of original reviews to be challenged (to a maximum of $\rho/2 + 1$ -- you cannot challenge all reviews). The editor solicits more reviews,  and if the challenge is upheld, the challenge cost is refunded to the author and deducted from the challenged reviewer(s).   This is intended to give more incentive to reviewers to be thorough.
    
    \item Editors may wish to give awards to particularly good reviewers, best papers, etc.  These extras are also folded into $\tau$.

    \item The amortized tax $\tau$ covers payments to every job in the conference: reviewers, area and vice chairs, program chairs, publicity chairs, etc.  This should ultimately make conferences \emph{more expensive} places to publish than, say journals that have far less organizational overhead.

    \item For the first couple of iterations, conferences should allow paper authors to borrow the submission cost against a promise to review for that conference. If they are unable to pay due to delinquent or low quality reviewing, their paper is withdrawn; this will leave an imbalance in the revenue since by the time the author is recognized as defaulting, their paper will have reviews that must be paid for, but no income for it.  I expect this will be a small fraction, however, and should be accounted for in $\tau$.

\end{itemize}

\subsection{Modeling the Tax Rate}
For simplicity, I recommend starting small and only paying the conference volunteers who handle papers, and adding other volunteers to the pay structure as the overall system becomes better understood.   

The volunteer pay structure should be a function of the number of papers each position handles, with the unit of work being 1 review = 1RC.  For example, the NeurIPS 2024 D\&B track had 2800 papers.  Each paper got 3-5 actual reviews and each assigned an AC, SAC, and TC, however the 3-5 reviews was the product of assigning 5 reviewers to each paper, expecting 40\% not to do it.  With review for pay and the stakes higher, we can expect that over-assignment to become uneccessary, and set $\rho=3$. The elements of $\tau$ are:

\begin{itemize}
    \item Track Chairs: 2800 papers, 0.125 RC/paper (split 3 ways)
    \item Senior Area Chairs: 100 papers, .25 RC/paper
    \item Area Chairs: 10 papers, .5 RC/paper
    \item extra reviews: 1/5 papers, .2 RC/paper
    \item loan defaults: (see above), .05 RC/paper
\end{itemize}

This simple logarithmic pay structure for the review hierarchy yields $\tau = 0.5 + 0.25 + 0.125 + 0.2 + 0.05 = 1.125 \approx 1$.  With $\rho=3$, this means the total outlay for the track would be $2800 \times 4 = 11,200$RC.

\section{Startup}
Clearly RC does not currently exist and in order to gather enough to submit, you need to review. Conferences will need to roll out their RC over a multi-year period.  Conferences can agree to share currencies, which will impact the modeling of the number of coins to mint, or different conferences could maintain their own coin.

\begin{itemize}
    \item Announce the new policy two years in advance of required payments.

    \item Determine the initial value of $\sigma$, the total number of RC coins to mint, based on analysis of previous conferences plus some padding for expected growth.  I expect the initial value of $\sigma$ to be roughly double the expected cost of running one conference: ie $\sigma = 2 \times n \times (\rho + \tau) $.  

    \item the value of $\tau$ (and therfore $\sigma$) requires setting payscales for all the conference volunteers and determing the number of extra reviews are needed per conference.  \textit{Prima facie} $\tau$ should be much less than $\rho$, as reviewing really is the primary thing a paper submitter is paying for.
    
    \item Disburse roughly $\sigma / 2$ RCs by paying the reviewers and volunteers from the most recent conference(s). 
    
    \item Disburse the rest of $\sigma$ (minting new coins if needed)  to the reviewers and volunteers from the next conference, for which which submission will be free.
\end{itemize}

This should leave enough coin around for the first conference to require payment, and will strongly incentivize legitimate reviewers to partipate in the last "free" conference so that they have RC to spend on the next one.

\section{Considerations}
This proposal was written with NeurIPS in mind, and covers the process as it operates as of Dec, 2024, though there is nothing particularly unusual about NeurIPS other than the fact that its current scale practically mandates some solution.  

The proposed system does not track papers, per se (e.g. by assigning them non-fungible tokens), but it does track researchers, as they need to pay with existing RC or be paid.   There have been documented cases of the same individual violating some conference submission policy and using an alias to bypass future restrictions, and in one purported case even reviewing their own paper.  This would not solve that problem completely, but it would be an improvement.

The proposed system does not rely on Openreview or ARR or other existing methods for tracking papers and reviews, though a much more sophisticated blockchain leger could be used.

It is certain that a secondary market for RC will emerge, and this should be considered a positive, as it increases the incentive for researchers to do reviews and take volunteer positions at conferences and journals. RC Scholarships and other kinds of charity can be added to the tax model, and should be strongly encouraged.

Long papers should receive special consideration for submission cost and payment.

With paper tracking systems like ARR, the (re)submission cost and payment for reviewing could be less.

Workshops and other satellite events could eventually adopt the same model, perhaps sharing the same RC.  Alternatively, or in addition, one could imagine particularly RC-rich researchers staking their own coin on putting together new events, instead of relying on rather difficult to assess workshop proposals.

Other forms of staking should be considered as well.

One can easily imagine organizations attempting to hoard RCs by requiring vulnerable members (employees, grad students) to transfer their RC to a central account. This is happening in some form or another already, e.g. companies that offer bounties for accepted papers, or companies able to perform certain higher profile research with costly infrastructure not widely available.  On the one hand, as long as the quality control on reviews is good, the community at least gets some benefit from the aggregation of resources. On the other, some monitoring of the hoarding of RC may be needed to be sure institutions aren't unfairly preventing others from joining the ecosystem.



\end{document}